\documentclass[onecolumn,11pt]{elsarticle}
\usepackage{epstopdf}
\usepackage{epsfig}
\usepackage{amssymb}
\usepackage{color}
\usepackage{bm}
\usepackage{amsfonts}
\usepackage{amsmath}
\usepackage{lineno,hyperref}
\usepackage{array}
\usepackage{microtype}
\usepackage[a4paper, total={6.25in, 8.5in}]{geometry}

\begin{document}
\title{A novel model of non-singular oscillating cosmology on flat Randall-Sundrum II braneworld}


		
\author{Rikpratik Sengupta$^a\footnote{$^*$Corresponding author.\\
				{\it E-mail addresses:} rikpratik.sengupta@gmail.com(RS)}$}
		
\address{$^a$Department of Physics, Aliah University, Kolkata 700 160, West Bengal, India}


\begin{abstract}
We obtain a \textit{novel} model of oscillating non-singular cosmology on the spatially flat Randall-Sundrum (RS) II brane. At early times, the universe is dominated by a scalar field with an inflationary emergent potential $V(\phi)=A(e^{B\phi}-1)^2$, $A$ and $B$ being constants. Interestingly, we find that such a scalar field can source a non-singular bounce, replacing the big bang on the brane. The turnaround again happens naturally on the brane dominated by a phantom dark energy (favoured by observations\cite{E1,E2,E3} at late times), thus avoiding the big rip singularity and leading upto the following non-singular bounce via a contraction phase. There is a smooth non-singular transition of the brane universe through both the bounce and turnaround, leading to alternate expanding and contracting phases. This is the \textit{first} model where a single braneworld of positive tension can be made to recycle as discussed in details in the concluding section. 
\end{abstract}

\maketitle



\section{Introduction}

The standard big bang model of cosmology is plagued by an initial singularity (at time $t=0$), where the Friedmann equations describing the time evolution of the universe fails to render a plausible physical description of the space-time dynamics due to the diverging Hubble parameter $H$, which is a consequence of the infinitely large energy density prevailing in the very early universe. The scalar curvature $R$ also diverges, implying that the initial singularity is a curvature singularity of the Ricci type, which are characterized by diverging energy densities. As shown in their famous singularity theorems by Hawking and Penrose, the initial big bang singularity cannot be avoided in a general relativistic (GR) setup provided the energy conditions are obeyed by matter filling up the universe\cite{Hawking,HP}. One of the key problems of the standard inflationary cosmology, besidies the ambiguity regarding the nature of the inflaton field is that, within the GR framework, inflation can not be past eternal\cite{Borde}. So, if the inflation is preceeded by a radiation dominated phase, then the origin of the universe is singular. However, today many cosmologists are unhappy about the initial singularity and consider it to be a limitation of GR to describe space-times involving very large energy densities.

If there is no big bang, then the possibilities are that the universe either experiences a quantum creation where there is a quantum mechanical tunelling into an inflationary phase, or it may be that the universe existed for an eternally long time in a quasi static state followed by an emergent inflationary phase, or there is a non-singular bounce replacing the singular big bang prior to which the universe contracts and following which the universe expands. In the context of GR, the emergent and bouncing scenarios can be realised effectively only for a spatially closed universe ($k=1$). A fully consistent treatment of the first possibility considering quantum creation will most probably require a quantum gravity (QG) treatment. However, there is no fully understood and developed QG theory at the moment and the two most accepted theories being worked on in this context are M-theory\cite{S2} involving extra dimensions and Loop Quantum Gravity (LQG)\cite{L1}. M- theory requires eleven space-time dimension for its quantum consistency while LQG quantizes space-time itself in the usual four dimensions. Effective theories from both the scenarios have become popular in recent times in the form of the extra dimensional braneworld models\cite{RS,SS} and the effective loop quantum cosmology (LQC)\cite{PS,PS2} models. 

It is interesting to note that although the premises of the background QG theories are completely different from each other, there are some similarities and identical features that may be obtained from a class of the effective braneworld and LQC models, that may hint at some hidden correspondance between the two contrasting approaches. In this letter, we shall only talk about models that introduce corrections to standard GR at the ultraviolet (UV) scale as we are concerned with resolution of the initial singularity. Braneworld models have the feature that our universe is represented by a $(3+1)$-dimensional hypersurface known as the 'brane' (which are objects appearing in M-theory) embedded in a higher dimensional bulk space-time. The Randall Sundrum single brane model (RS-II) is one such model with a spacelike extra dimension, indicating that the signature of the bulk space is Lorentzian. If the signature of the bulk deviates from the former, then we can have a bulk signature $(-,-,+,+,+)$ such that the braneworld has a timelike extra dimension. The standard model particles and fields are confined to the brane, while gravity is free to propagate in the bulk.

\section{The cosmological model}

The Einstein Field Equations (EFE) on the brane have the general form

\begin{equation}
		G_{\mu \nu}+\Lambda h_{\mu \nu}=8\pi G_{eff} T_{\mu \nu}+\frac{\epsilon}{1+\beta}\bigg[\frac{S_{\mu \nu}}{M^6}-W_{\mu \nu}\bigg]
\end{equation}

The parameter $\epsilon=\pm1$, where the + sign implies a spacelike and - sign a timelike extra dimension. $G_{\mu \nu}$ and $T_{\mu \nu}$ denote the Einstein and stress-energy tensors on the brane, respectively while $h_{\mu\nu}$ is the induced metric on the brane. We have $8\pi G_{eff}=\frac{\beta}{\beta+1}\frac{1}{m^2}$ and $\Lambda=\frac{\Lambda_{br}}{\beta+1}$, where the parameter $\beta=\frac{2 \epsilon \sigma m^2}{3M^6}$ and $\Lambda_{br}=\frac{\Lambda_{bulk}}{2}+\frac{\epsilon \sigma^2}{3M^6}$. Here $G_{eff}$ is the effective Gravitational constant on the brane, $\Lambda_{br}$ and $\Lambda_{bulk}$ denotes the effective cosmological constant on the brane and the bulk cosmological constant, respectively. $m$ and $M$ denote the Planck masses in $4$ and $5$ dimensions, respectively and $\sigma$ represents the brane tension. In the RS limit, we have $m\rightarrow 0$ implying $\beta \rightarrow 0$, such that the effective gravitational constant is given by $G_{eff}=\frac{\epsilon \sigma}{12\pi M^6}$. The sign of $\epsilon$ can also be interpreted as the sign of $M$ or alternatively as the sign of the brane tension as either $M$ or $\sigma$ has to be negative for $\epsilon=-1$ to make the gravitational constant positive. Thus, a braneworld with a timelike extra dimension is characterized by a negative brane tension in order to localize the bulk gravity in the vicinty of the brane. Also, $\Lambda_{bulk}>0$ for such a brane, as contrasted to the RS II brane which is characterized by an anti-deSitter ($AdS_{5}$) bulk. To sum up, in the RS limit, the modified EFE is written with $\Lambda$ replaced by $\Lambda_{br}$, $8\pi G_{eff}$ replaced by $\frac{2\epsilon \sigma}{3M^6}$ and $\beta=0$. Finally, coming to the tensors $S_{\mu \nu}$ and $W_{\mu \nu}$, they represent the local and non-local corrections on the brane, where the former represents quadratic corrections in stress-energy components given as $S_{\mu \nu} =  \frac{T T_{\mu \nu}}{12}  - \frac{ T_{\mu \alpha} T^{\alpha}_{\nu}}{4}  + \frac{g_{\mu \nu} }{24} \left[ 3 T_{\alpha \beta} T_{\alpha \beta} - (T^{\alpha}_{\alpha})^{2} \right]$ and the later represents the projected bulk Weyl tensor on the brane given by $W_{\mu \nu}=C_{ABCD}^{bulk}n^C n^D h_{\mu}^A h_{\nu}^B$, where $C_{ABCD}^{bulk}$ is the bulk Weyl tensor and $n^C$ is unit vector normal to the surface.

For the FRW line element describing an isotropic, homogenous universe, the modified EFE on the brane has the form
\begin{equation}
	H^2+\frac{k}{a^2}=\frac{\Lambda_{bulk}}{6}+\frac{\epsilon \sigma^2}{9M^6}+\frac{2 \epsilon \sigma \rho}{9M^6}+\frac{\epsilon^2 \rho^2}{9M^6}+\frac{C}{a^4}.
\end{equation} 
   
Observations suggest a spatially flat universe\cite{Efstathiou,SV1,Sv2,SV3} enabling us to put $k=0$ and also we can consider $C=0$ due to the high symmetry of the space-time to obtain a simpler solution, without much loss of generality. RS also considered the effective cosmological constant on the brane ($\Lambda_{br}$) to be $zero$ as the bulk cosmological constant and the parameter $\epsilon$ have opposite signs and so the first two terms on the right hand side of Equation (3) ($=\frac{\Lambda_{br}}{3}$) can be fine tuned to vanish. After some algebraic simplification, the modified EFE on the brane takes the form
\begin{equation}
	H^2= \frac{8\pi G_{eff}}{3}\rho\bigg(1+\frac{\epsilon \rho}{2 |\sigma|}\bigg)
\end{equation}	

Coincidentally, Equation (3) for a timelike extra dimension ($\epsilon=-1$) gives the same modified EFE as obtained from the LQC perspective, although the frameworks are inspired from completely different background theories, as discussed\cite{SS,PS}. The factor involving the brane tension $|\sigma|$ in the denomenator of the second term on the RHS acts as a critical density $\rho_{cr}$ in LQC. A very interesting feature of such models is that they can remove the initial singularity which can be understood very simply from Equation (3) as follows: the energy density cannot rise beyond the critical density as for $\rho=\rho_{cr}=2|\sigma|$, the Hubble parameter vanishes. So, the big bang singularity is replaced by a non-singular big bounce even in a spatially flat universe, which is not possible in standard GR. This can be attributed to some repulsive force that is generated as the energy density increases, solely due to the extra dimension being timelike or due to the quantization of spacetime. The identical effect and result may indicate a deep underlying connection between the different mechanisms.

However, in this letter we shall try to construct a ${novel}$ non-singular cosmological model of oscillating nature on the spatially flat RS II brane with a spacelike extra dimension ($\epsilon=1$). For a spacelike extra dimension $\epsilon$ being positive, the Hubble parameter does not vanish naturally if the energy density $\rho$ rises upto 2$\sigma$. So, RS II cosmology does not contain any inherent feature through which a non-singular bounce can be naturally realized. A Weyl curvature singularity due to infinitely large tidal forces at the wormhole throat can be resolved on the RS II brane with ordinary matter\cite{RS1}. We shall try to induce the bounce using an ingredient that is quite familiar in early universe cosmology and finds application in achieving the inflationary mechanism- a scalar field. The scalar field is minimally coupled meaning that there is no coupling between the scalar field and gravity. In such a model there may exist an infinite number of cycles containing expansion and contraction phases. However, another additional mechanism is required besides a non-singular bounce to generate such an oscillating universe. The universe must begin to contract at late times following an expanding phase before the next cycle begins. This mechanism is known as the turnaround. We shall use two different mechanisms to generate the bounce and turnaround on the flat RS II brane.       

The conservation equation on the brane is the same as in GR given by $\dot{\rho}+3H(p+\rho)=0$, where for the scalar field $\phi$, the energy density is expressed as $\rho=\frac{\dot{\phi}^2}{2}+V(\phi)$ and the pressure has the form $p=\frac{\dot{\phi}^2}{2}-V(\phi)$. Here $V(\phi)$ represents the potential of the scalar field. So, the conservation for the scalar field on the brane has a form identical to standard GR and is given as
\begin{equation}
	\ddot \phi + 3H\dot\phi + \frac{dV}{d\phi} = 0
\end{equation}

We shall consider a scalar field of potential having a flat wing from which an inflationary emergent scenario can be realized in GR. Such a well known potential having the form $V(\phi)=A(e^{B\phi}-1)^2$ was used by Ellis et al.\cite{Ellis} to construct an emergent universe scenario in the relativistic setting for a spatially closed universe. The potential as a function of the field has been plotted in Figure 1, where $A$ and $B$ denote constant parameters. It can be considered that the scalar field is weakly decelerated ($|\ddot{\phi}|<|\frac{dV}{d\phi}|$) as it is the left branch of the potential that can possibly lead to a non-singular cosmology. Under this consideration, the $\ddot{\phi}$ in the equation of motion for the scalar field is of negligible significance. 

Making use of the modified Friedmann equation and the conservation equation for a weakly decelerating scalar field of potential $V(\phi)$ having the form as mentioned above, a solution for the field $\phi(t)$ is obtained to be given by
\begin{equation}
	\phi(t) = \frac{1}{3}\frac{2B^2\sqrt{6\sigma}t + 3BC_1\kappa + 3LambertW\left(\frac{2B^2\sqrt{6\sigma}t - 3BC_1\kappa}{3\kappa}-1\right)\kappa}{B\kappa},
\end{equation} 
where $\kappa=\sqrt{8 \pi G_{eff}}$.

\begin{figure}[!htp]
	\centering
	\includegraphics[width=5cm]{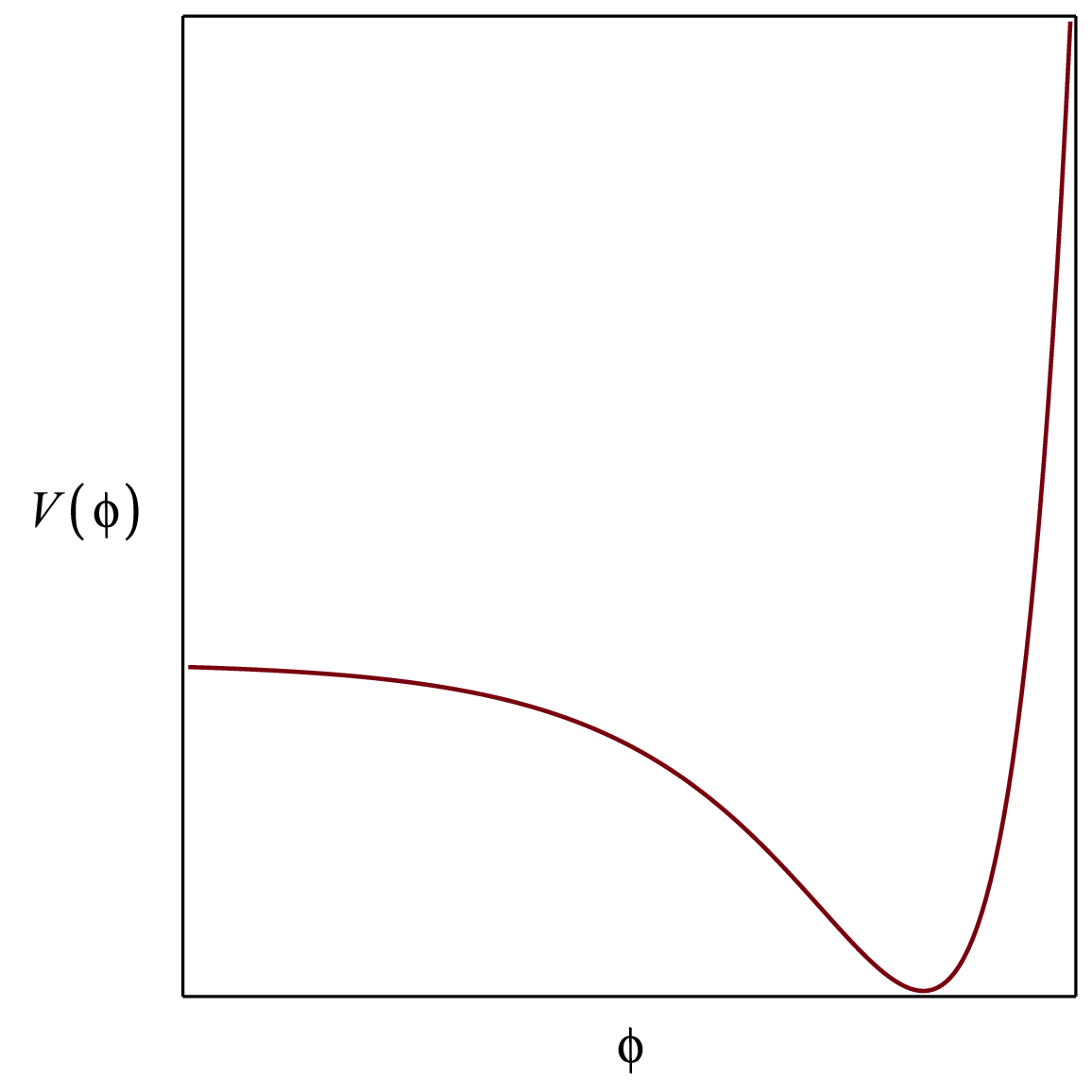}
	\includegraphics[width=5cm]{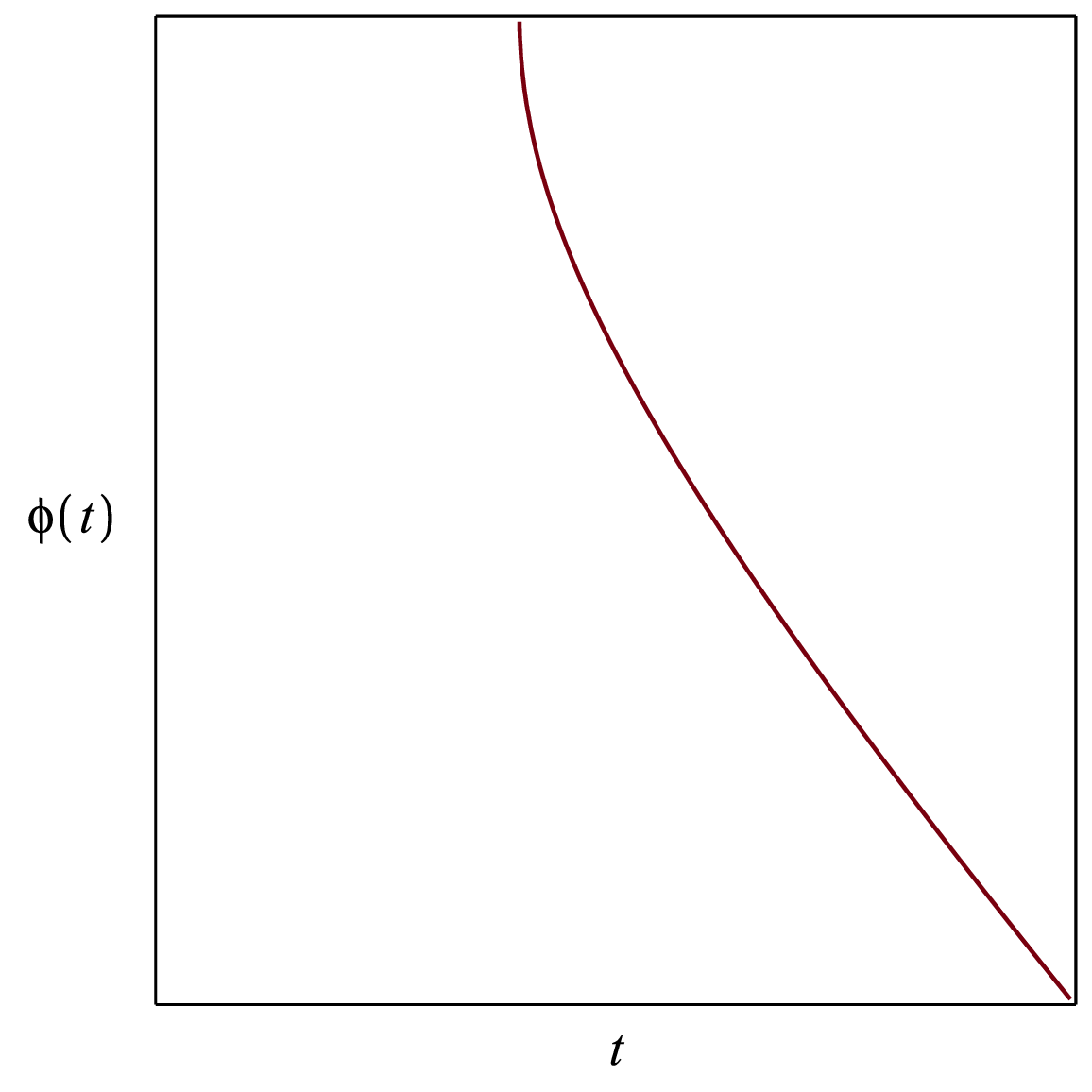}
	\caption{Variation of the scalar field potential $V(\phi)$ with the field ($\phi$)and the field $\phi(t)$ with time $t$.}
\end{figure} 
The time evolution of the scalar field has also been plotted in Figure 1. As evident from the figure (estimating the time of the bounce from Figure 2), negative values of the field play an important role in the bounce. So, from the evolution of the potential, it may be said that the bounce happens for the flatter branch of the potential that leads to an emergent cosmology in a closed relativistic Friedmann universe.

We know that the Hubble parameter is defined as $H=\frac{\dot{a}}{a}$ and the scale factor $a$ can be expressed in tetrm of the scalar field $\phi(t)$ as
\begin{equation}
	a(\phi) = C_2e^{-\frac{1}{4}\frac{A \kappa^2}{B\sigma}\left[\frac{e^{2B \phi}}{2B} - \phi - 3\frac{e^{B \phi}}{B} + \frac{e^{-B \phi}}{B}\right]},
	\label{scale_first_pot_RS}
\end{equation}
where $C_2$ is a constant of integration.

Using Equations (5) and (6), we have
\begin{multline}
	a(t) = C_2e^{-\frac{1}{4}\frac{A\kappa^2}{B\sigma}}\left[\frac{e^{\frac{4B^2\sqrt{6\sigma}t + 3BC_1\kappa + 3LambertW\left(\frac{2B^2\sqrt{6\sigma}t - 3BC_1\kappa}{3\kappa} -1\right)\kappa}{3\kappa}}}{2B}\right.
	\\ \left.-\frac{1}{3}\frac{2B^2\sqrt{6\sigma}t + 3BC_1\kappa + 3LambertW\left(\frac{2B^2\sqrt{6\sigma}t - 3BC_1\kappa}{3\kappa}-1\right)\kappa}{B\kappa}\right.
	\\ \left. -3\frac{e^{\frac{2B^2\sqrt{6\sigma}t + 3BC_1\kappa + 3LambertW\left(\frac{2B^2\sqrt{6\sigma}t - 3BC_1\kappa}{3\kappa}-1\right)\kappa}{3\kappa}}}{B} + \frac{e^{-\frac{2B^2\sqrt{6\sigma}t + 3BC_1\kappa + 3LambertW\left(\frac{2B^2\sqrt{6\sigma}t - 3BC_1\kappa}{3\kappa}-1\right)\kappa}{3\kappa}}}{B} \right]
	\label{scale_first_pot_RS_time}
\end{multline} 
 
\begin{figure}[!htp]
	\centering
	\includegraphics[width=5cm]{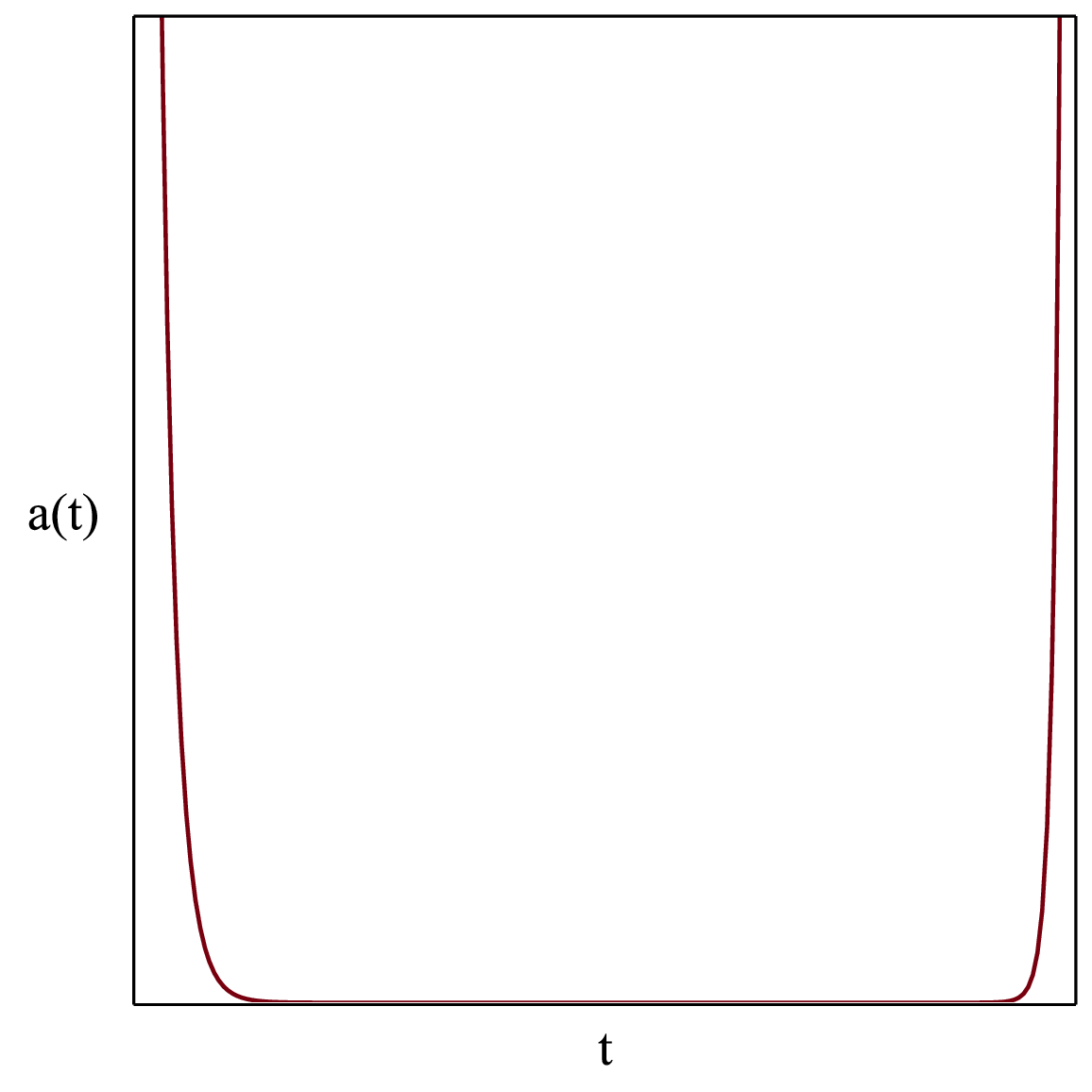}
	\includegraphics[width=5cm]{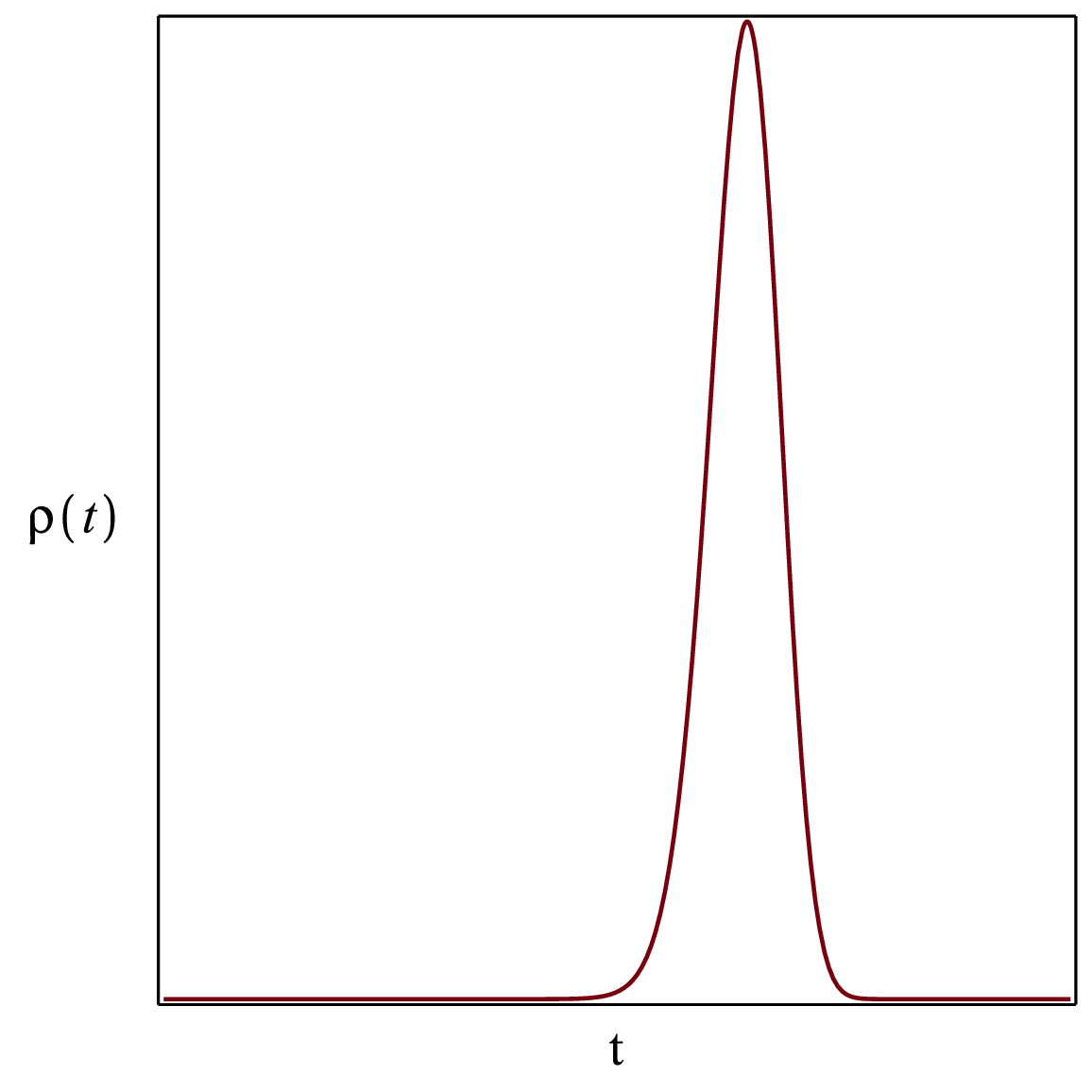}
	\caption{Variation of the scale factor $a(t)$ and the energy denisty $\rho(t)$ with time $t$ showing the non-singular bounce.}
\end{figure}

As we can see from the variation of the scale factor with time in Figure 2, a scalar field with the potential of the form $V(\phi)=A(e^{B\phi}-1)^2$ can cause the brane universe to bounce without reaching a singular state. The contracting universe bounces into an expanding phase before a singular state of diverging energy density and scalar curvature is reached. The varaiation of the energy density of the scalar field dominated universe with time has also been shown in Figure 2 around the time of the bounce. At the bounce, the energy density reaches a maximum as expected, making the $\rho^2$ term in the UV corrected EFE significant, before falling off again following it. The effectiveness of the scalar field in averting the initial singularity and generating the bounce can be realized through this quadratic energy density term reflecting the higher dimensional effect due to the braneworld scenario. We shall not go into the details of whether there is a subsequent inflationary phase to seed cosmological perturbations, although it is known that inflation can be obtained from both the branches (flat and steep) of the potential that we have used. We shall focus only on the beginning of the cycle through the bounce and its completion through a turnaround leading to the beginning of the next cycle. The detailed evolution of the cycle including the generation of large scale structures from seed perturbations is beyond the scope of this letter and shall be taken up in a following work. 

We now move on to the late time where the universe is in an accelerating phase. Such a phase has been inferred from astronomical observations\cite{P1,R1}. With the discovery of the accelerating universe, there was a resurrection of the $\Lambda$-term in cosmology. However, there are certain inconsistencies with the $\Lambda$ dark energy (DE) that has lead to a wide range of models including quintessence involving scalar fields\cite{De1,De4}, Chaplygin gas involving fluids with non-linear equation of states\cite{Cg1,Cg2} (EoS), the phantom with an exotic EoS\cite{Pe1,Pe3} and geometrical models which effectively modify the matter sector through geometrical contributions at the infrared (IR) scale\cite{DGP,DDG,NojiriR} but not the real source matter. Here, we have already used a framework in which the geometrical contributions modify the matter sector effectively at the UV scale through the $\rho^2$ term. If such a term has to be significant in the universe at late times to source a turnaround through its possible non-conventional behaviour ($H\propto \rho$), then there has to be a mechanism to make the energy density of the universe grow sufficiently large enough. There is a possibility to achieve this using one of the possible DE candidates that is observationally well favoured-the phantom. The phantom is an exotic fluid with a supernegative EoS ($\omega<-1$) violating the Null Energy condition (NEC), proposed by Caldwell\cite{Pe1} to fit the observational data. The fact tha the phantom fits quite well with the observational data ($-1.61<\omega<-0.78$) has latter been verified by a number of groups\cite{E1,E2,E3}.       

We have already discussed that the conservation equation on the brane is the same as in GR. For a perfect fluid described the the EoS $p=\omega \rho$, the conservation equation can be used to obtain the evolution of the energy density in terms of the scale factor as $\rho \simeq a^{-3(1+\omega)}$. As for the phantom $\omega<-1$, so the factor in the power of $a$ is always a positive number, resulting in $\rho(t)$ growing with $a$ for a phantom dominated universe. This makes the UV corrected $\rho^2$ term significant also at late times. In a relativistic DE scenario with a phantom fluid, there appears a few bizzare features. If we consider $t'$ to be the time when the density of ordinary matter dominating the contacting phase of the universe is equal to that of the phantom DE, before DE domination (accelerating expansion) takes over, then the scale factor during the later phase evolves as $a(t)=a(t')\bigg[(1+\omega)\frac{t}{t'}-\omega\bigg]^{\frac{2}{3(1+\omega)}}$, which  blows up to infinity within finite time $t=\frac{\omega t'}{1+\omega}$, making the universe singular. This singularity is termed as the big rip. At the same finite time, the energy density of the phantom dominated universe $\rho(t) \simeq \bigg[(1+\omega)\frac{t}{t'}-\omega\bigg]^{-2}$ also blows up to infinity making the Hubble parameter diverge. The curvature scalar also diverges resulting in a curvature singularity like the big bang. However, as we shall see, the quadratic correction term in the modified EFE can avoid this singularity too, by making the universe turnaround and enter a contracting phase before the energy density diverges, resulting in the next non-singular bounce dominated by a scalar field with the potential discussed.   

Again making use of the UV corrected Friedmann equation on the brane, we obtain
\begin{equation}
	\int \frac{da}{\bigg[a^{\frac{-1-3\omega}{2}}+\frac{\alpha}{2\sigma}a^{\frac{-4-6\omega}{2}}\bigg]}=\sqrt{\frac{\kappa^2 \alpha}{3}}\int dt,
\end{equation} 
where $\alpha$ is a constant.

We obtain analytical solutions for three different values of the EoS parameter, considering a phantom dominated universe.

\begin{eqnarray}
	&&(\omega=-\frac{4}{3}): a(t)=\frac{1}{\bigg[LambertW(\frac{1}{2}+2t^2)+1\bigg]^2}   \nonumber\\
	&&(\omega=-\frac{3}{2}): a(t)=\frac{1}{\bigg[LambertW(\frac{1}{2}+\frac{9}{8}t^2)+1\bigg]^\frac{4}{3}}  \nonumber\\
	&&(\omega=-3): a(t)=\frac{1}{\bigg[LambertW(\frac{1}{2}+6t^2)+1\bigg]^\frac{1}{3}}  
\end{eqnarray}

\begin{figure}[!htp]
	\centering
	\includegraphics[width=5cm]{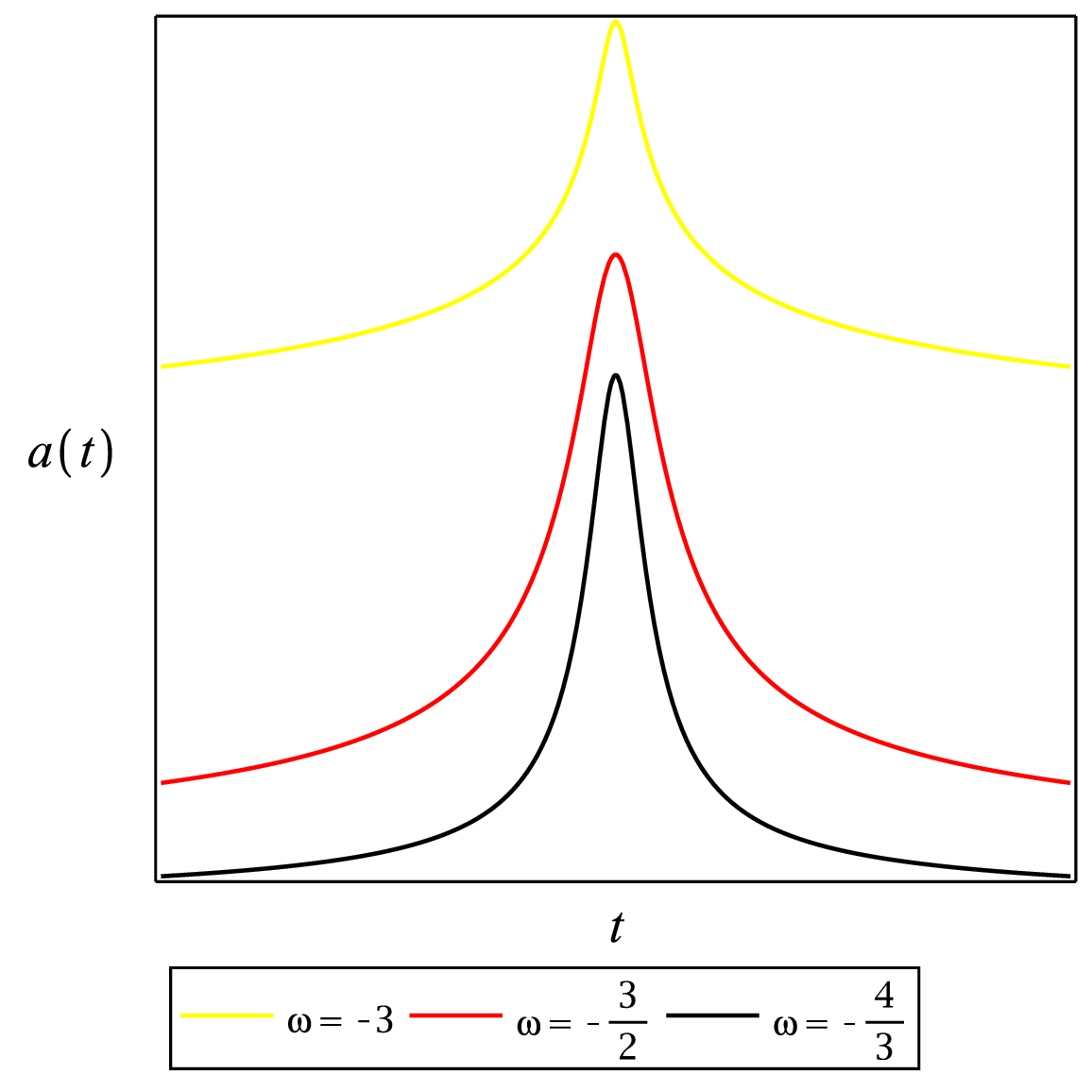}
	\includegraphics[width=5cm]{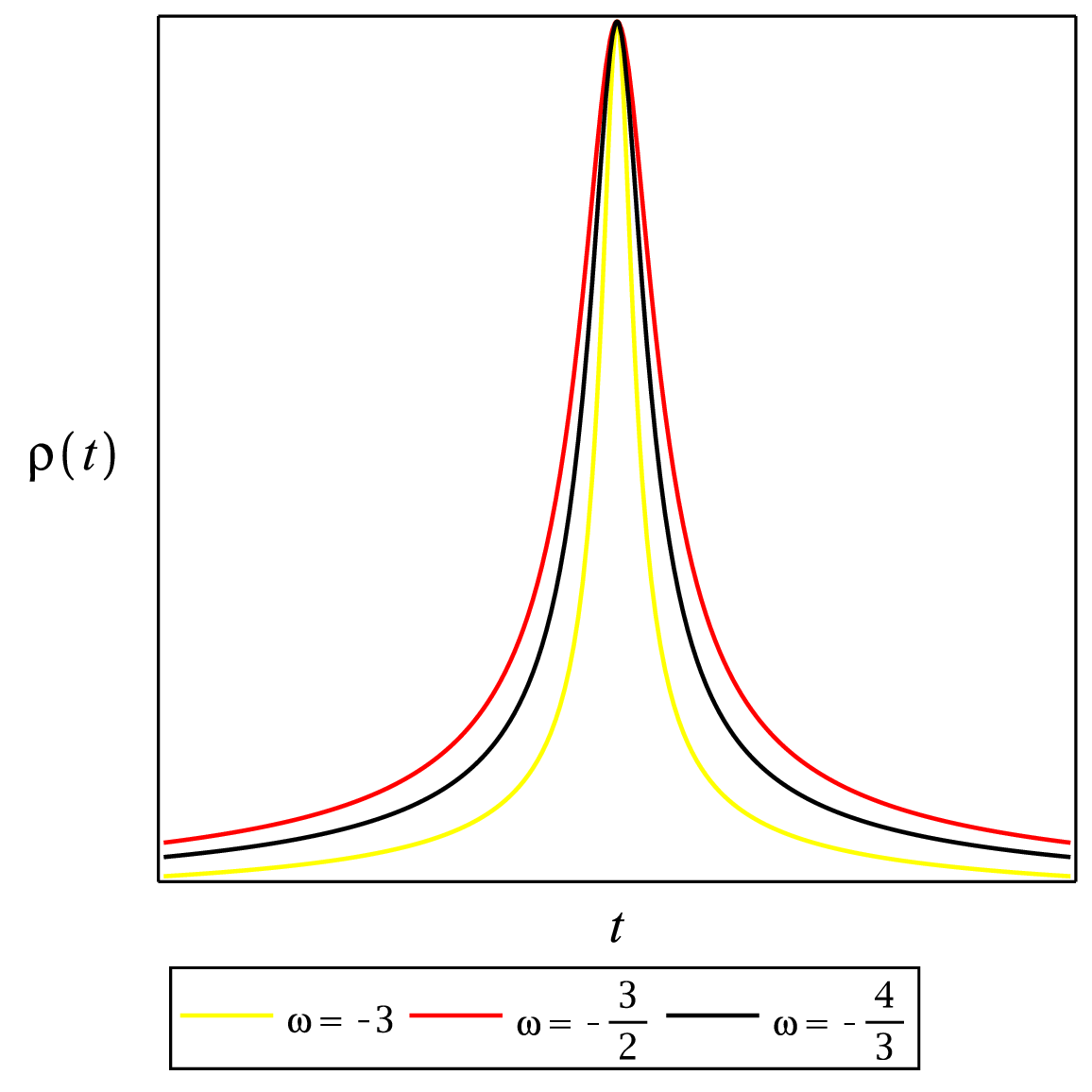}
	\caption{Variation of the scale factor $a(t)$ and the energy denisty $\rho(t)$ with time $t$ showing the turnaround for different values EoS parameter $\omega$ of the dominant cosmic fluid.} 
\end{figure}

As we find from Figure 3 on plotiing the scale factor with time, the accelerating phantom dominated universe enters into a contracting phase before the scale factor diverges. Once the universe begins to contract, it is dominated by radiation or matter until the scalar field begins to dominate the universe that causes it to bounce before the scale factor reaches a zero value. So, the scale factor maintains a non zero finite value all along the evolution of the universe, never reaching a singular state. The energy density of the universe has also been plotted in Figure 3 as it evolves with time close to the turnaround. It is found to reach a peak just before the turnaround as it keeps on increasing in the phantom dominated eoch, before falling off again. The energy density is large enough at both early and late times to make the quadratic correction term in the modified EFE significant, but never diverges. It starts dropping following both the bounce and turnaround.  

\section{Discussion}

The two most important components of our oscillating cosmological model on the brane are the bounce and turnaround mechanisms. The bounce mechanism is sourced by a scalar field with the inflationary emergent potential $V(\phi)=A(e^{B\phi-1})^2$ that may be followed by an inflationary phase (or some other mechanism leading to an increase in amplitude of each successive cycle) to seed cosmological perturbations resulting in the formation of large scale structures in the universe. This shall be a subject of future concern. However, if there is an inflationary phase, it can be sourced from either of the two branches (flat and steep) of the potential mentioned above for suitable values of the parameters $A$ and $B$. However, the left branch leads to an emergent cosmology and also to a bouncing cosmolgy as obtained by us. The bounce is possible due to the \textit{effective violation of NEC by the scalar field on the brane.}  So, no additional source may be required to seed perturbations in the early universe. The turnaround mechanism is sourced by the phantom energy described a supernegative EoS ($\omega<-1$). Such a component can not only explain the accelerating phase of the universe but also source the turnaround due to \textit{the unique property that the energy density grows with time making the UV correction term significant also at late times, that ultimately leads to the turnaround on the brane}. So, no additional component is required for this purpose also. Moreover, we obtain the cyclic scenario for an observationally \textit{favoured} spatiall flat universe\cite{SV1,SV2,SV3}.

When the universe is contracting, the energy density grows and ultimately diverges making the scalar curvature and the Hubble parameter to diverge as well. This can be understood by the fact that the scale factor vanishes in the Friedmann equation. Bounce is the mechanism that simply prevents the \textit{initial} singularity from forming by making the scale factor begin to increase before it can reach zero or making the energy density drop off before it can diverge. The condition for bounce is achieved through $\ddot{a}>0$, such that the contracting universe starts expanding. Alternatively, in turnaround, the expanding universe must begin to start contracting so that both the scale factor and energy density donot diverge in the finite future and this can be achieved through the condition $\ddot{a}<0$. Both at the bounce and turnaround, the \textit{Hubble parameter vanishes} rather than diverging. The the scale factor neither reaches zero or infinite value as the \textit{effective energy density on the brane remains finite.} Thus, the universe transits smoothly through both the bounce and turnaround. For cosmology on the brane, both the mechanisms can be achieved by a minimum of requisite components- a scalar field with an inflationary potential and a DE component that \textit{violates the NEC}.

The problem with the phantom is that such an exotic fluid has a number of theoretical inconsitencies and pathologies at the quantum level that makes their existence questionable. The problem with the future singularities can be cured from the correction term on the brane as we have found, but such a fluid may also lead to the vacuum being unstable. Attempts to construct dynamical scalar field models of the phantom have led to a negative kinetic term\cite{Pe1} that in turn results in quantum instabilities\cite{Moore}. However, there exists a cosmological model of DE with vanishing $\Lambda$ where the vacuum energy obtained from the quantization of a free scalar field having low mass is described by a supernegative EoS and the model is free of pathologies at the quantum level\cite{Parker}. Two problems in general feature in most oscillating cosmology models. We shall discuss them very briefly without going into much details before concluding the letter. The first problem is posed by the continued existence of singular objects like black holes from the area theorems of Hawking. However, before the turnaround leading up to the next bounce (via a contraction phase) in a phantom dominated universe, such structures may well be dissolved due to the extremely large gravitationally repulsive effects\cite{Caldwell}, thus being prevented from disrupting the evolution of the universe during the contracting phase following the turnaround. Infact, it has been shown that\cite{Davies} the Hawking area theorems may not hold true if the NEC ($\rho+p\geq 0$) is violated, as is the case for a phantom dominated universe. Any surviving remnant microscopic black holes may act as possible dark matter candidates. Moreover, the black hole singularity may also be resolved in the UV corrected picture just like the initial big bang and the big rip singularities and also, there may exist non-singular black hole mimickers like gravastar on the brane\cite{RS2}, leading to a complete resolution of the problem at once. It is worth mentioning in this context that the RSII braneworld has also been used in explaining a recent GW event GW170817\cite{Visinelli} and the recent observation of the dark shadow of M87$^{\ast}$\cite{Vagnozzi}. The second problem is associated with the entropy of the universe, which we think remains the same in a periodic manner after the bounce over each cycle, such that the possible increase in entropy during the expanding phase being compensated by a possible decreasing during the radiation/matter dominated expanding phase. This prevents the enrtopy to increase to infinitely large values limiting the number of cycles. We are however concerned mainly with the bounce and turnaround in a single cycle in this letter.    

We have constructed a \textit{novel} model of non-singular oscillating cosmology on the RSII brane where the universe in a contracting phase bounces into expansion sourced by a scalar field before the energy density diverges, thus resolving the big bang singularity and is taken over by a phantom fluid at the late times (causing the universe to accelerate as observed and with an increasing energy density) making the $\rho^2$ term in the EFE on the brane contribute, thus avoiding a big rip singularity from forming in the finite future (via the universe entering into a contracting phase which continues but again re-enters into an expanding phase, due to the non-singular bounce occuring in an oscillating manner). In between the scalar field and phantom fluid is the standard radiation and matter dominated epochs. Thus, there are alternate expanding and contracting epochs.      

This is \textit{the first model that can avert the initial singularity using a single brane approach with a positive brane tension.} Braneworlds that have a space-like extra dimension like the one we have considered here are characterized by a positive brane tension (as the effective gravitational constant on the brane needs to be positive to explain the attractive nature of gravity) but such a setup could not resolve the big bang singularity. The non-singular models of brane cosmology till date have either resorted to a single brane with time-like extra dimension where no scalar field has to be invoked to generate the bounce which happens naturally from the cosmological dynamics\cite{SS}, but the brane tension must be negative for the same reason of obtaining a positive effective gravitational constant, or alternatively,  introduce a second braneworld with a negative tension parallel to the positive tension brane with finite separation between the branes. The advantage of introducing the parallel negative tension braneworld is two-fold: firstly, negative tension branes have the unique feature of reduced inertia on matter with positive energy density being dumped onto it helping the dynamical realization, and secondly the two brane setup comes with the benefit of a scalar field known as the radion which modulates the inter-brane separation and can both source the bounce at early times as well as behave like phantom dark energy at late times dure to the non-canonical kinetic term evolving to have a negative value\cite{SRS}. However, there are some tachyonic instabilities associated with negative tension braneworlds which can be possibly resolved in M-theory but has not been explored well enough and requires further formal developments in M-theory (although the properties are really appealing). On the contrary, the ingredients of our model are well explored and the physics is more well understood in terms of a single positive tension brane. The phantom dark energy also does not lead up to the big rip as the quadratic correction to stress energy becomes significant before the singularity can is reached.  
		
Also, the scalar field we have used is a physically well motivated one, since it can accommodate the inflationary scenario naturally and its potential need not be reconstructed to explain the generation of seed cosmological perturbations. Most models of cosmology with a non-singular bounce have either to resort to alternative mechanisms to generate seed perturbations that are not very well understood physically, or have to reconstruct  the potential  on an ad hoc basis to generate the perturbations, but in our model an inflationary epoch following the bounce driven by the scalar field with an inflationary emergent potential already has all the ingredients responsible for generating these perturbations, and is well understood. We may think of it as a toy model, not because the scenario is physically ill-motivated, but since we have not tested the model against observations. We plan to analyze the primordial observables like the amplitude of scalar perturbations, tensor to scalar ratio and the spectral index and test them against latest observations in a follow up work in the recent future.

\section*{Acknowledgments}
RS is thankful to the Inter-University Centre for Astronomy and Astrophysics (IUCAA),
Pune, India for providing work facilities during a visit where this work was done. 

\textbf{Data Availability Statement} This is a theoretical work and does not make use of any data for arriving at our results. However, the observation supporting our analysis have been cited in each case in the manuscript.

\end{document}